# Symmetric and Asymmetric Quantum Rabi Model

Moorad Alexanian

*Department of Physics and Physical Oceanography*
*University of North Carolina Wilmington, Wilmington, NC 28403-5606*

E-mail: alexanian@uncw.edu

**Abstract.** We introduce a modified Jaynes-Cummings model with single-photon cavity radiation field but with the atomic system instead of exchanging a single photon as in the Jaynes-Cummings model, it exchanges instead a squeezed photon. After a unitary transformation and requiring the decoupling of the spin up from the spin down, we diagonalize the resulting Hamiltonian via a Bogoliubov transformation. This allows to determine the energy eigenvalues for the quantum Rabi model. We obtain the energy eigenvalues albeit for the asymmetric Rabi model in the limit of large coupling strength, where it approaches the symmetric Rabi model energy eigenvalues.

## 1. Introduction

The quantum Rabi model describes the interaction between a two-level quantum system (qubit) and a single-mode quantum electromagnetic field and is quite useful in quantum optics and solid-state settings [1]. A review gives exact analytic solutions in terms of confluent Heun functions and an overview of several experimental realizations of the quantum Rabi model [2]. A quantum phase transition in the quantum Rabi model has been observed with a single trapped ion [3]. The present paper is motivated by two previous papers [4, 5], which method and assumptions we strictly follow.

In this paper, we introduce a modified Jaynes-Cummings model [4], where the radiation field is given by a single-mode photon while the atomic system exchanges squeezed photons instead of single photons. This model encompasses both the Jaynes-Cummings model as well as the quantum Rabi model as limiting cases. The present paper is arranged as follows. In Sec. 2, the modified Jaynes-Cummings model is introduced. In Sec. 3, the energy eigenvalue spectrum is determined together with the expression for the photon number operator. In Sec. 4, we treat the asymmetric Rabi model in the limit of large coupling strength between the two-level system and the bosonic mode. Finally, in Sec. 5 we summarize our results.

## 2. Modified Jaynes-Cummings model

In a recent paper [4], a modified Jaynes-Cummings model was introduced, where the atomic transition was via the exchange of a squeezed photon rather than the exchange of a photon, with the resulting Hamiltonian

$$\hat{H}_{MJC} = \hbar\omega_c \hat{a}^\dagger \hat{a} + \hbar\omega_a \frac{\hat{\sigma}_z}{2} + \frac{\hbar\Omega}{2}(\hat{\sigma}_+ \hat{B} + \hat{\sigma}_- \hat{B}^\dagger), \qquad (1)$$

where $\omega_a = \omega_2 - \omega_1$, with $\hbar\omega_1$, $\hbar\omega_2$ are the energies of the uncoupled states $|1\rangle$ and $|2\rangle$, respectively, and $\omega_c$ is the frequency of the field mode. The system can be in two possible states $|i\rangle$, $i = 1, 2$ with $|1\rangle$ being the ground state of the system and $|2\rangle$ being the excited state, respectively. The annihilation operator for the squeezed photon $\hat{B}$ is

$$\hat{B} = \cosh(r)\hat{a} + e^{i\varphi}\sinh(r)\hat{a}^\dagger. \qquad (2)$$

Hamiltonian (1) can be expressed as

$$\hat{H}_{MJC} = \hbar\omega_c \hat{a}^\dagger \hat{a} + \hbar\omega_a \frac{\hat{\sigma}_z}{2} + \frac{\hbar\Omega}{2}\left[\cosh(r)(\hat{\sigma}_+\hat{a} + \hat{\sigma}_-\hat{a}^\dagger) + \sinh(r)(\hat{\sigma}_-\hat{a} + \hat{\sigma}_+\hat{a}^\dagger)\right], \qquad (3)$$

where we have chosen $\varphi = 0$.

After a unitary transformation $\hat{U} = e^{-v(\hat{\sigma}_+\hat{B}^\dagger - \hat{\sigma}_-\hat{B})}$ that decouples terms between spin subspaces $\mathcal{H}_\downarrow$ and $\mathcal{H}_\uparrow$ since $\langle\downarrow|\sigma_\pm|\downarrow\rangle = 0$, one has

$$\hat{H} \equiv \langle\downarrow|\hat{U}^\dagger \hat{H}_{MJC}\hat{U}|\downarrow\rangle = A\hat{a}^\dagger\hat{a} + B + C(\hat{a} + \hat{a}^\dagger)^2, \qquad (4)$$

where

$$A = \hbar\omega_c(1 + v^2) + \hbar v(\Omega + \omega_a v)e^{-2r}, \qquad (5)$$

$$B = -\frac{\hbar\omega_a}{2}(1 - v^2) + \frac{\hbar v}{2}(\Omega + \omega_a v)e^{-2r} + \hbar\omega_c v^2 \cosh^2(r), \qquad (6)$$

and

$$C = -\frac{\hbar v\Omega}{2}\cosh(2r) + \frac{\hbar\omega_a v^2}{2}\sinh(2r). \qquad (7)$$

The Hamiltonian (4) is diagonalized with the aid of the linear Bogoliubov transformation

$$\hat{a} = \cosh(\beta)\hat{b} + \sinh(\beta)\hat{b}^\dagger$$
$$\hat{a}^\dagger = \cosh(\beta)\hat{b}^\dagger + \sinh(\beta)\hat{b}. \qquad (8)$$

The cancellation of the terms $\hat{b}\hat{b} + \hat{b}^\dagger\hat{b}^\dagger$ implies that

$$A\cosh(\beta)\sinh(\beta) + C\left[\cosh(\beta) + \sinh(\beta)\right]^2 = 0. \qquad (9)$$

Therefore,

$$\hat{H} = B - A/2 + \sqrt{A(A + 4C)}\;(\hat{b}^\dagger\hat{b} + 1/2), \qquad (10)$$

where the constants $A$, $B$, $C$ are given by (5), (6), (7), respectively.

## 3. Symmetric Rabi model

One obtains the symmetric Rabi model from Hamiltonian (3) in the limit $r \to \infty$ such as $\hbar\Omega e^r/4 \to g$ and so

$$\hat{H}_{SR} = \hbar\omega_c \hat{a}^\dagger\hat{a} + \hbar\omega_a \frac{\hat{\sigma}_z}{2} + g(\hat{a} + \hat{a}^\dagger)\hat{\sigma}_x, \qquad (11)$$

since In addition, $ve^r/2 \to k$ for a finite unitary transformation

The above limiting processes give rise to the following values for *A*, *B*, *C* in (5)-(7)

$$\hat{\sigma}_+ + \hat{\sigma}_- = \hat{\sigma}_x.$$

$$\hat{U} \to e^{-ik(\hat{a}+\hat{a}^\dagger)\hat{\sigma}_y}.$$

$$\begin{aligned} A &= \hbar\omega_c \\ B &= -\hbar\omega_a/2 + \hbar\omega_c k^2 \\ C &= -2kg + \hbar\omega_a k^2 \end{aligned} \qquad (12)$$

Quantum phase transitions take place at absolute zero, viz., the ground state, where a bifurcation into two phases occurs. The quantum phase transition is characterized by $\epsilon \equiv \sqrt{A(A+4C)} = 0$, which determines the critical values $k_\pm$ associated with the parameter *k* in the unitary transformation $\hat{U}$ and so

$$\epsilon = 4g_c\sqrt{(k-k_+)(k-k_-)}, \qquad (13)$$

where

$$k_\pm = \frac{g}{\hbar\omega_a}\left(1 \pm \sqrt{1 - \frac{g_c^2}{g^2}}\right), \qquad (14)$$

provided $kg > 0$, where $g_c$ is the critical coupling strength for the normal/superradiance transition in both the Rabi and the Jaynes-Cummings models [4], viz.,

$$g_c^2 = \lambda_c^2 = \frac{\omega_a \omega_c \hbar^2}{4}. \qquad (15)$$

If we choose $k = g/\hbar\omega_a$, then Hamiltonian (10) becomes [5]

$$\hat{H} = -\frac{\hbar\omega_a}{2}\left(1 + \frac{\omega_c}{\omega_a}\right) + \frac{\omega_c}{\hbar\omega_a^2}g^2 + \hbar\omega_c\sqrt{1 - \frac{g^2}{g_c^2}}\left(\hat{b}^\dagger\hat{b} + \frac{1}{2}\right). \qquad (16)$$

In addition, if one considers the limit $\omega_a/\omega_c \to \infty$, which corresponds to the thermodynamic limit in the Rabi model, then our result (16) reduces to the result $H_{np}$ for the normal phase (np) [5].

The photon number can be obtained by setting $\omega_a = 0$ and $\Omega = 0$ in (3), (5)-(7) and with the aid of (10)

$$\hat{n} \equiv \langle\downarrow|\hat{U}^\dagger \hat{a}^\dagger \hat{a}\, \hat{U}|\downarrow\rangle = v^2\cosh^2(r) + (1+v^2)\hat{b}^\dagger\hat{b}, \qquad (17)$$

which in the limit $v \to 2ke^{-r}$ becomes

$$\hat{n} = \hat{b}^\dagger\hat{b} + k^2. \qquad (18)$$

Note that the photon number operator $\hat{a}^\dagger\hat{a}$ and $\hat{b}^\dagger\hat{b}$ possess the same eigenstates, with eigenvalues *n* and *m*, respectively, and so

$$n = k^2 + m, \qquad (19)$$

where $m = 0, 1, 2, \cdots$ and $n = 1, 2, \cdots$ with cavity photon number eigenstates exist only provided $k^2$ is a nonnegative integer, viz., the quantization of the unitary parameter $k = \sqrt{l}$, with $l = 1, 2, \cdots$. It is important to remark that for fixed $m$, $n \to \infty$ as $k^2 \to \infty$. Note that the converse is not true, that is, for fixed $n$, the value of $k^2$ is bounded from above, viz., $n \geq k^2$.

The energy E

$$E_m(n) = -\frac{\hbar}{2}(\omega_a + \omega_c) + \hbar\omega_c k^2 + \sqrt{\hbar\omega_c(\hbar\omega_c - 8kg + 4\hbar\omega_a k^2)}\,(m+\frac{1}{2}), \quad (20)$$

where $k$ is given by (19). In what follows, we shall analytically continue $k^2$ to non-integer values, viz., $k^2$ a continuous variable. The asymptotic behavior for $n \to \infty$ for given $m$ follows from (19)-(20)

$$E_m(n) = \hbar\omega_c n + 2\hbar\sqrt{\omega_a\omega_c}(m+1/2)\sqrt{n} - \frac{\hbar}{2}(\omega_a+\omega_c) - \hbar\omega_c m - 2g\sqrt{\frac{\omega_c}{\omega_a}}(m+\frac{1}{2}) + O(\frac{1}{\sqrt{n}}). \quad (21)$$

This asymptotic behavior differs from results obtained by de Monvel and Zielinski [6] but in agreement with the limit of the level spacing

$$E_m(n+1) - E_m(n) \to \hbar\omega_c \qquad n \to \infty. \quad (22)$$

It is curious that the number operator for the cavity photons (18) is only related to the unitary parameter $k$ and not related to the coupling strength $g$, which are both truly independent variables. However, at the quantum phase transition, they are indeed related by (13) hence

$$\hat{n}_\pm = \hat{b}^\dagger\hat{b} + \frac{g^2}{\omega_a^2\hbar^2}\left[1 \pm \sqrt{1 - \frac{g_c^2}{g^2}}\right]^2. \quad (23)$$

Fig.1 shows the behavior of the modified unitary parameter $\kappa_\pm \equiv 2\sqrt{\omega_a/\omega_c}\,k_\pm$ for both branches of the bifurcation with $k_\pm$ given by (14). Fig.2 shows the phase diagram demarcating the normal phase (np) from the superradiance phase (sp). Fig.3 shows the asymptotes to the quantum phase transition curve, viz., $g/g_c = (\kappa^2+1)/(2\kappa)$, , that is, $g/g_c = \kappa/2$ (green) and $g/g_c = 1/(2\kappa)$ (red).

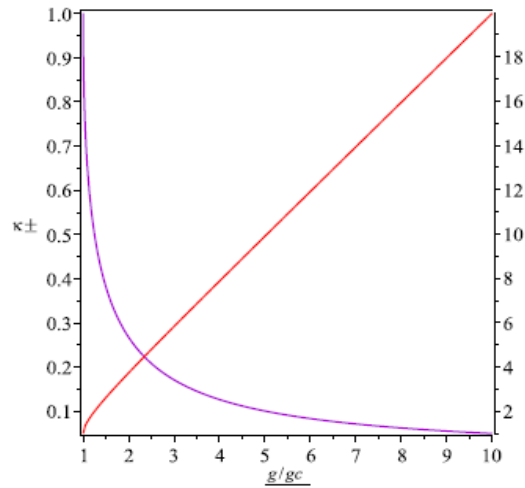

**Fig. 1.** Phase change diagram $\kappa_\pm \equiv 2\sqrt{\omega_a/\omega_c}\,k_\pm$ vs. reduced coupling strength $g/g_c$ at the quantum phase transition $\epsilon \equiv \sqrt{A(A+4C)} = 0$. The upper branch $\kappa_+$ (red) while the lower branch $\kappa_-$ (violet) from the bifurcation (14).

One obtains the following expression for the energy along the asymptote $g/g_c = \kappa/2$ in the normal phase

$$\frac{E(m)}{\hbar\omega_c} + \frac{1}{2}\frac{\omega_a}{\omega_c} = m + k^2 = \text{integer}, \quad (24)$$

Similarly, along the asymptote $g/g_c = 1/(2\kappa)$,

$$2\frac{g}{g_c}\left(\frac{E(m)}{\hbar\omega_c} + \frac{1}{2}\frac{\omega_a}{\omega_c} + \frac{1}{2} - \frac{1}{64}\left(\frac{\hbar\omega_c}{g}\right)^2 - \frac{1}{2}\frac{g_c}{g}\right) = m = \text{integer}, \tag{25}$$

also in the normal phase. Note that the two asymptotes intersect at $k = 1$ giving $g/g_c = 1/2$.

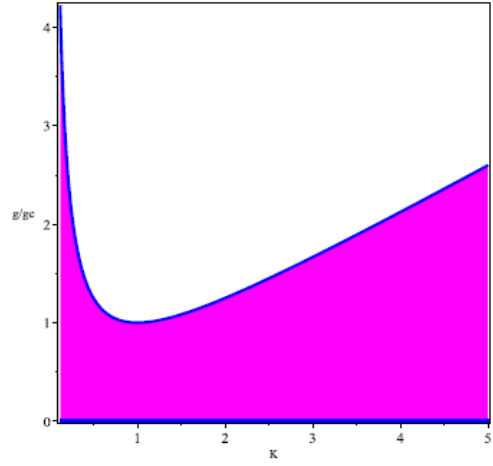

**Fig. 2.** Universal reduced coupling strength $g/g_c$ at the quantum phase transition $\epsilon \equiv \sqrt{A(A+4C)} = 0$ (blue), viz., the boundary of the phase transition $g/g_c = (\kappa^2 + 1)/(2\kappa)$, where $\kappa = 2k\sqrt{\omega_a/\omega_c}$. The upper region is the superradiance phase (sp) (blank), where $\epsilon$ is purely imaginary, while the lower region (magenta), where $\epsilon$ is real, is the normal phase (np).

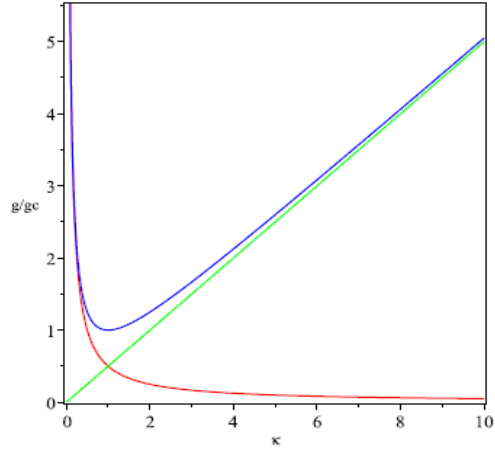

**Fig. 3.** Asymptotes to phase transition curve $g/g_c = (\kappa^2 + 1)/(2\kappa)$ (blue), where $\kappa = 2k\sqrt{\omega_a/\omega_c}$, see Fig. 2. Asymptotes $g/g_c = \kappa/2$ (green), $g/g_c = 1/(2\kappa)$ (red).

## 4. Asymmetric Rabi model

Consider the following modification of the symmetric Rabi model (11)

$$\hat{H}_{MSR} = \hat{D}(-\alpha)\hat{H}_{SR}\hat{D}(\alpha), \tag{26}$$

where the Glauber displacement operator is

$$\hat{D}(\alpha) = \exp(\alpha\hat{a}^\dagger - \alpha^*\hat{a}) \tag{27}$$

and so

$$\hat{H}_{MSR} = \hbar\omega_c(\hat{a}^\dagger + \alpha^*)(\hat{a} + \alpha) + \hbar\omega_a \frac{\hat{\sigma}_z}{2} + g(\hat{a} + \alpha + \hat{a}^\dagger + \alpha^*)\hat{\sigma}_x \qquad (28)$$

One obtains, in the limit $\alpha \to 0$ and $g \to \infty$ such that $g(\alpha + \alpha^*) \to \epsilon$, the asymmetric Rabi model albeit for large values of $g$

$$\hat{H}_{AR} = \hbar\omega_c \hat{a}^\dagger \hat{a} + \hbar\omega_a \frac{\hat{\sigma}_z}{2} + g(\hat{a} + \hat{a}^\dagger)\hat{\sigma}_x + \epsilon\hat{\sigma}_x. \qquad (29)$$

Accordingly, the spectrum of the asymmetric Rabi model is the same as that of the symmetric Rabi model albeit for large values of $g$. In the normal phase, the energy $\epsilon$ in (13) is real and so

$$\frac{\kappa^2 + 1}{2\kappa} \geq \frac{g}{g_c}, \qquad (30)$$

where $\kappa = 2k\sqrt{\omega_a/\omega_c}$. Therefore, there are two possible behaviors for $g \to \infty$, viz., $\kappa \to \infty$ or $\kappa \to 0$. For instance, along the asymptote $g/g_c = \kappa/2$, that is, $k = 2g/(\hbar\omega_a)$, which gives the maximal increasing $g$ in the normal phase, we obtain for the energy (20) that

$$\frac{E(m)}{\hbar\omega_c} = m + \frac{4g^2}{\hbar^2\omega_a^2} - \frac{\omega_a}{2\omega_c}. \qquad (31)$$

Note that along this asymptote, large values of $g$ correspond to large values for the integer $l$ since $k = \sqrt{l}$, $l = 1, 2, 3\cdots$ and so $k$ becomes actually a continuous variable. Note also that in a recent work, the energy actually decreases with increasing values of $g$ [7]. The limit $g \to \infty$ implies, for fixed $m$, that $n \to \infty$ according to (19).

On the other hand, along the asymptote $g/g_c = 1/(2\kappa)$, that is, $k = \hbar\omega_c/(8g)$. We have from (20) that

$$\frac{E(m)}{\hbar\omega_c} = -\frac{1}{2} - \frac{\omega_a}{2\omega_c} + \frac{\hbar^2\omega_c^2}{64g^2} + \frac{g_c}{2g}\left(m + \frac{1}{2}\right). \qquad (32)$$

The limit $g \to \infty$ implies $k \to 0$ and so $n - m \to 0$, that is, the number of photons approaches that of the quasiparticles (19). Note that along the asymptotes in the limit $g \to \infty$, the energy is independent of $\epsilon$ as expected.

## 5. Conclusions

A modified Jaynes-Cummings model whereby the atomic transition is mediated by a squeezed photon gives rise to a quantum normal/superradiance phase transition. The phase diagram $g - k$, between the coupling strength $g$ and the parameter $k$ associated with a unitary transformation, delineates the normal from the superradiance phases. We obtain the energy of the symmetric quantum Rabi model for all values of $g$ and $k$ and the energy of the asymmetric quantum Rabi model for large values of $g$.